\documentstyle[twocolumn,aps,prl]{revtex}

\makeatletter

\providecommand{\LyX}{L\kern-.1667em\lower.25em\hbox{Y}\kern-.125emX\@}

\makeatletter

\newcommand{\mathbold}[1]{\mbox{\boldmath $\bf#1$}}
\makeatother

\makeatother

\begin{document}

\wideabs{
 \title{Damping of low-energy excitations of a Bose-condensed gas in the hydrodynamic
regime.}
 \author{P.O. Fedichev\protect\( ^{1,2}\protect \), A.E. Muryshev\protect\( ^{2}\protect \)
and G.V. Shlyapnikov\protect\( ^{1,2}\protect \)}
 \address{\protect\( ^{1}\protect \)FOM Institute for Atomic and Molecular Physics,
Kruislaan 407, 1098 SJ Amsterdam, The Netherlands.}
 \address{\protect\( ^{2}\protect \)Russian Research Center, Kurchatov Institute, Kurchatov
Square, 123182 Moscow, Russia.}
 \maketitle
\begin{abstract}
We develop a theory to describe the damping of elementary excitations of a Bose-condensed gas 
in the hydrodynamic regime for the thermal cloud. We discuss second sound in a spatially 
homogeneous gas and the lowest excitations of a trapped condensate.
\end{abstract}
}

The discovery of Bose-Einstein condensation (BEC) in trapped alkali atoms vapors
\cite{Cor95,Ket95,Hul95} has stimulated a growing activity in the field of
ultra-cold gases \cite{Pitaevskii:review}. Recent developments allow one to
reach sufficiently high densities and temperatures of a Bose-condensed gas,
where the thermal cloud leaves the collisionless domain and enters the collision
dominated (hydrodynamic) regime \cite{MIT:shifts}. This is the regime that
has been studied in superfluid liquid helium. However, due to the high density
and the complexity of interparticle interaction in helium, only few results
can be fully described on the basis of an \emph{ab initio} theory. In this respect,
the study of hydrodynamic excitations of a Bose-condensed gas is important,
since it bridges together phenomenological models of condensed
matter physics and microscopic theories.

The damping of elementary excitations of a Bose-condensed gas in the hydrodynamic
regime is qualitatively different from that in the collisionless regime. In
the latter case the excitation frequency \( \omega  \) greatly exceeds
a characteristic relaxation rate in the thermal cloud \( \tau_R^{-1} \), and this
cloud behaves as an ensemble of Bogolyubov quasiparticles.
Theories of damping phenomena for a Bose-condensed gas in the collisionless
regime are reviewed in \cite{Pitaevskii:review} and for the earlier work in
\cite{PopovBook}.

In the hydrodynamic regime, where \( \omega \tau _{R}\ll 1 \), 
the thermal cloud
rapidly responds to the motion of the condensate. Accordingly, quasiparticles
of the cloud loose their individual character, and the spectrum changes. Then
the dynamics of a Bose-condensed gas is described by the Landau-Khalatnikov
(LKh) two-fluid hydrodynamics \cite{khalatnikov:book} characterized by two
independent branches of elementary excitations: first and second sound. 
Recently a microscopic derivation of the two-fluid hydrodynamics
has been generalized for the case of trapped gases (see \cite{griffin:review}
and refs. therein).
For a homogeneous Bose-condensed gas a microscopic theory equivalent to the LKh
hydrodynamics was developed by Popov \cite{popov:hydro}. Thus far this theory
has only been used for calculating transport coefficients at very low temperatures
\cite{PopovBook}.

In this Letter we develop a theory to describe the damping of low-energy
excitations of a Bose-condensed gas in the high-temperature hydrodynamic regime.
We discuss second sound in a spatially homogeneous gas and the lowest excitations of a trapped 
condensate at temperatures \( T\gg \mu  \) (the chemical potential \( \mu =gn_{0} \),
with \( n_{0} \) being the condensate density, \( g=4\pi \hbar ^{2}a/m \),
\( m \) the atom mass, and \( a \) the scattering length) and assume that the density of
non-condensed atoms \( n^{\prime}\ll n_0 \). 
In this case, second sound representing out-of-phase oscillations of the condensate
and thermal component, is mostly seen as the condensate oscillations, since the amplitude of 
oscillations of the thermal component is much smaller (see \cite{GZ}). 
The friction between the condensate and the normal component, originating from
collisions between Bose-condensed and thermal particles, leads to energy transfer
from the condensate to the thermal cloud and, hence, to damping of the condensate oscillations. 

We will start with calculating the damping of second sound in a 
homogeneous gas.
At \( T\gg \mu  \) and \( n^{\prime}\ll n_0 \) 
the velocity of second sound is \( c_S=\sqrt{n_0g} \) (hereafter \( \hbar=m=1 \)) and it
is much smaller than the velocity of first sound \( u\sim \sqrt{T} \) \cite{popov:hydro}
representing oscillations of the thermal component. In this sense, the two branches of the excitation
spectrum are adiabatically separated. 
For these conditions, the deviation of the condensate
density \( \delta n_0(t,{\bf r}) \) from its equilibrium value \( n_{0} \) satisfies the hydrodynamic equations
\cite{LL:volIX}
\begin{equation}
\label{hydrolin}
\frac{\partial \delta n_{0}(t,{\textbf {r}})}{\partial t}+n_{0}\mathbold {\nabla }{\textbf {v}}=0,\; \; m\frac{\partial {\textbf {v}}}{\partial t}+g\mathbold {\nabla }\delta n_{0}(t,{\textbf {r}})=0,
\end{equation}
 where \( {\textbf {v}} \) is the superfluid velocity. The solutions of Eq.(\ref{hydrolin})
are sound waves with momentum \( {\textbf {k}} \) and frequency \( \omega _{{\textbf {k}}} \)
satisfying the dispersion relation \( \omega _{k}=c_{S}k \).

As we will see below, the damping of the condensate oscillations is mostly provided
by thermal excitations with energies \( \epsilon \sim T\gg \mu  \). In a dilute
gas they are always in the collisionless regime and, hence, the excitation energy
is \( \epsilon _{p}=p^{2}/2+gn_{0}(t,{\textbf {r}}), \) where \( {\textbf {p}} \)
is the momentum of the excitation, and \( n_{0}(t,{\textbf {r}})=n_{0}+\delta n_{0}({\textbf {r}},t) \)
is the slowly oscillating condensate density. 
We will assume that these thermal excitations are quasiclassical and
describe them by using a classical distribution function \( f(t,{\textbf {r}},{\textbf {p}}) \).

The behavior of the thermal excitations is governed by the Hamiltonian \( H_{0}+H_{{\textrm{int}}} \),
where 
\begin{eqnarray}
H_{0} & = & \int d^{3}r\sum _{{\textbf {p}}}\left( \frac{p^{2}}{2}+gn_{0}\right) f(t,{\textbf {r}},{\textbf {p}}),\label{H0} \\
H_{int} & = & \int d^{3}r\sum _{{\textbf {p}}}g\delta n_{0}(t,{\textbf {r}})f(t,{\textbf {r}},{\textbf {p}}).\label{Hint} 
\end{eqnarray}
 Oscillations of the condensate density can be written as 
\begin{equation}
\label{deltan}
\delta n_{0}(t,{\textbf {r}})=i\sum _{{\textbf {k}}}\left( \frac{\omega _{k}}{2n_{0}g}\right) ^{1/2}(c_{{\textbf {k}}}e^{i{\textbf {kr}}-i\omega _{k}t}-c.c.),
\end{equation}
 with \( c_{{\textbf {k}}} \)(\( c^{*}_{{\textbf {k}}} \)) being a classical
limit of the creation(annihilation) operators of second sound phonons.
It is the interaction term \( H_{{\textrm{int}}} \)(\ref{Hint}) that leads
to energy transfer from the condensate oscillations (phonons of second sound)
to the thermal cloud.

We will perform our analysis within a linear response theory. In the presence
of phonons, the equilibrium distribution function of thermal excitations \( f_{0}({\textbf {p}})=(\exp {(\epsilon _{p}/T)}-1)^{-1} \)
acquires a small extra term \( \delta f \). The Fourier transforms of \( \delta f(t,{\textbf {r}},{\textbf {p}}) \)
and \( \delta n_{0}(t,{\textbf {r}}) \), 
\[
\left( \begin{array}{c}
\delta f_{{\textbf {p}}}(\omega ,{\textbf {k}})\\
\delta n_{0}(\omega ,{\bf k})
\end{array}\right) =\int dtd{\textbf {r}}\exp {(i{\textbf {kr}}-i\omega t)}\left( \begin{array}{c}
\delta f(t,{\textbf {r}},{\textbf {p}})\\
\delta n_{0}(t,{\bf r})
\end{array}\right) ,\]
 are related to each other by means of a generalized susceptibility \( \alpha _{{\textbf {p}}}(\omega ,{\textbf {k}}) \)
\cite{com1}: 
\begin{equation}
\label{f}
\delta f_{{\textbf {p}}}(\omega ,{\textbf {k}})=\alpha _{{\textbf {p}}}(\omega ,{\textbf {k}})(\partial f_{0}/\partial \epsilon _{p})g\delta n_{0}(\omega ,{\textbf {k}}).
\end{equation}

The imaginary part of \( \alpha _{{\textbf {p}}}(\omega ,{\textbf {k}}) \)
is responsible for dissipation and is related to the damping rate \( \Gamma _{k} \)
of the (first sound) phonon with momentum \( {\textbf {k}} \) and frequency
\( \omega _{k} \): 
\begin{equation}
\label{losses:delta}
\Gamma _{k}=\omega _{k}g{\textrm{Im}}\langle \langle \alpha _{{\textbf {p}}}(\omega _{k},{\textbf {k}})\rangle \rangle .
\end{equation}
Eq.(\ref{losses:delta}) follows from the fact that the energy loss rate of the condensate oscillations is
given by \( \sum _{{\textbf {k}}}\omega _{k}|c_{{\textbf {k}}}|^{2}\Gamma _{k} \)
and should be equal to the rate of the energy increase in the thermal cloud.
The latter reduces to \( \dot{H}=\sum _{{\textbf {k}}}\omega _{k}|g\delta n_{0}(\omega ,{\textbf {k}})|^{2}{\textrm{Im}}\langle \langle \alpha _{{\textbf {p}}}(\omega ,{\textbf {k}})\rangle \rangle  \),
where \( \langle \langle ...\rangle \rangle \equiv \sum _{{\textbf {p}}}\partial f_{0}/\partial \epsilon _{p}(...) \)
(see \cite{pitaevskii:resonances}). 

The response function \( \alpha _{{\textbf {p}}} \) can be found from the solution
of the kinetic equation for the thermal excitations. As the force acting on
a thermal particle is \( -g\mathbold {\nabla }n_{0} \), the kinetic equation
takes the form 
\begin{equation}
\label{kineq:original}
\frac{\partial f}{\partial t}+{\textbf {p}}\frac{\partial f}{\partial {\textbf {r}}}-g\mathbold {\nabla }n_{0}\frac{\partial f}{\partial {\textbf {p}}}=I_{{\textbf {p}}}[f],
\end{equation}
 where \( I_{{\textbf {p}}}[f] \) is the collisional integral. There are two
types of binary collisions contributing to \( I_{{\textbf {p}}}[f] \): collisions
of non-condensed with Bose-condensed atoms, transferring the latter to the
thermal cloud (and vice versa), and collisions between non-condensed atoms,
leaving both of them in the thermal cloud. The former process is characterized
by the relaxation rate \( \tau _{R}^{-1}\sim n_{0}\sigma v_{T} \), where \( \sigma =8\pi a^{2} \)
is the elastic cross section and \( v_{T} \) the thermal velocity, and the
latter process by the rate \( n'\sigma v_{T} \).
Assuming the inequality \( n'\ll n_{0} \), we
confine ourselves only to collisions involving Bose-condensed atoms. The corresponding
collisional integral is 
\[
I_{{\textbf {p}}_{1}}[f]=2\pi g^{2}n_{0}\sum _{2,3}[(1/2)\delta (\epsilon _{{\textbf {p}}_{1}}-\epsilon _{{\textbf {p}}_{2}}-\epsilon _{{\textbf {p}}_{3}})\delta _{{\textbf {p}}_{1},{\textbf {p}}_{2}+{\textbf {p}}_{3}}\times \]
 
\[
((1+f_{1})f_{2}f_{3}-f_{1}(1+f_{2})(1+f_{3}))+\delta (\epsilon _{{\textbf {p}}_{1}}-\epsilon _{{\textbf {p}}_{2}}+\epsilon _{{\textbf {p}}_{3}})\times \]
 
\[
\delta _{{\textbf {p}}_{1},{\textbf {p}}_{2}-{\textbf {p}}_{3}}((1+f_{1})(1+f_{3})f_{2}-(1+f_{2})f_{1}f_{3})],\]
 where \( f_{i}\equiv f(t,{\textbf {r}}_{i},{\textbf {p}}_{i}) \), and the
factor \( 1/2 \) in front of the first term in the rhs appears due to the double
counting of final states. In the spatially homogeneous case the condition \( n'\ll n_{0} \)
requires temperatures well below the BEC transition temperature \( T_{c} \),
whereas in a trapped gas it is satisfied even at \( T \) close to \( T_{c} \)
(see, e.g., \cite{Pitaevskii:review}). Since in the linear regime the deviation
\( \delta f \) of the distribution function from the equilibrium value \( f_{0} \)
is small, we linearize the collisional integral \( I_{{\textbf {p}}} \). Introducing
the function \( h_{{\textbf {p}}}(t,{\textbf {r}})=\delta f(t,{\textbf {r}},{\textbf {p}})(\partial f_{0}/\partial \epsilon _{p})^{-1} \)
and setting \( I_{{\textbf {p}}}[f_{0}+h\partial f_{0}/\partial \epsilon _{p}]=-(\partial f_{0}/\partial \epsilon _{p})\tilde{I}_{{\textbf {p}}}[h] \),
we obtain 
\[
\tilde{I}_{{\textbf {p}}_{1}}[h]=\pi g^{2}n_{0}(1-\exp {(-\epsilon _{{\textbf {p}}_{1}}/T)})\sum _{2,3}f_{0}(\epsilon _{{\textbf {p}}_{2}})f_{0}(\epsilon _{{\textbf {p}}_{3}})\times \]
 
\[
(\exp {(\epsilon _{{\textbf {p}}_{1}}/T)}\delta (\epsilon _{{\textbf {p}}_{1}}\! \! -\! \epsilon _{{\textbf {p}}_{2}}\! -\epsilon _{{\textbf {p}}_{3}})\delta _{{\textbf {p}}_{1},{\textbf {p}}_{2}+{\textbf {p}}_{3}}(h_{{\textbf {p}}_{1}}\! -h_{{\textbf {p}}_{2}}\! -h_{{\textbf {p}}_{3}})+\]
\begin{equation}
\label{colintlinear}
\! \! 2\exp \! {(\epsilon _{{\textbf {p}}_{2}}/T)}\delta (\epsilon _{{\textbf {p}}_{1}}\! \! \! -\! \epsilon _{{\textbf {p}}_{2}}\! \! \! +\! \epsilon _{{\textbf {p}}_{3}})\delta _{{\textbf {p}}_{1},{\textbf {p}}_{2}\! \! -\! {\textbf {p}}_{3}}(h_{{\textbf {p}}_{1}}\! \! -\! h_{{\textbf {p}}_{2}}\! \! +\! h_{{\textbf {p}}_{3}})).\! \! 
\end{equation}

The collisional integral has a number of important properties \cite{LL:volX}.
Collisions conserve the total momentum and energy of particles, and thus \( \tilde{I}[{\textbf {p}}]={\textbf {0}} \)
and \( \tilde{I}[\epsilon _{p}]=0 \). In addition, the collisional integral
is orthogonal to the integrals of motion: For an arbitrary function \( \chi  \)
we have 
\begin{equation}
\label{orthog:short}
\langle \langle (\begin{array}{c}
{\textbf {p}}\\
\epsilon _{p}
\end{array})\tilde{I}_{{\textbf {p}}}[\chi ]\rangle \rangle =(\begin{array}{c}
{\textbf {0}}\\
0
\end{array}).
\end{equation}

We now use Eq.(\ref{f}) and rewrite the kinetic equation (\ref{kineq:original})
in the Fourier representation: 
\begin{equation}
\label{kineq:alpha}
L[\alpha _{{\textbf {p}}}]\equiv (\omega -{\textbf {kp}})\alpha _{{\textbf {p}}}+{\textbf {kp}}=-i\tilde{I}_{{\textbf {p}}}[\alpha _{{\textbf {p}}}].
\end{equation}
 Eqs. (\ref{kineq:alpha}) and (\ref{losses:delta}) allow us to find the response
of the thermal cloud to the condensate oscillations and to calculate the damping
rate for the modes of second sound. In the hydrodynamic regime (\( \omega \tau _{R}\ll 1 \)),
Eq.(\ref{kineq:alpha}) is dominated by the collisional integral and can be
solved by using the Chapman-Enskog method (we are following the Popov ansatz
for a Bose-condensed gas \cite{popov:hydro}). We will seek the solution of
Eq.(\ref{kineq:alpha}) in the form of expansion in powers of a small hydrodynamic
parameter \( \omega \tau _{R} \): 
\begin{equation}
\label{response:generalform}
\alpha _{{\textbf {p}}}=\alpha ^{(0)}_{{\textbf {p}}}+\alpha _{{\textbf {p}}}^{(1)}+\alpha _{{\textbf {p}}}^{(2)}+...,
\end{equation}
 where \( \alpha ^{(i+1)}_{{\textbf {p}}}/\alpha ^{(i)}_{{\textbf {p}}}\sim \omega \tau _{R}\ll 1 \).
The collisional integral can be estimated as \( \tilde{I}[\alpha ]\sim \alpha /\tau _{R} \)
and, hence, we have \( \tilde{I}[\alpha _{{\textbf {p}}}^{(i+1)}]\sim \omega \alpha _{{\textbf {p}}}^{(i)} \).
Therefore, Eq.(\ref{kineq:alpha}) is equivalent to an infinite set of coupled
equations: 
\begin{eqnarray}
\tilde{I}_{{\textbf {p}}}[\alpha _{{\textbf {p}}}^{(0)}] & = & 0,\label{ladder1} \\
L[\alpha _{{\textbf {p}}}^{(i)}] & = & -i\tilde{I}_{{\textbf {p}}}[\alpha ^{(i+1)}_{{\textbf {p}}}].\label{ladder2} 
\end{eqnarray}

In the so called acoustic approximation, one neglects all terms with \( i>0 \)
in Eq.(\ref{response:generalform}). The zero order solution \( \alpha _{{\textbf {p}}}^{(0)} \)
follows from Eq.(\ref{ladder1}) and is a linear combination of the integrals
of motion: 
\begin{equation}
\label{alpha:zero}
\alpha _{{\textbf {p}}}^{(0)}=a\epsilon _{p}+b{\textbf {kp}}.
\end{equation}
 The functions \( a \) and \( b \) should be found from the substitution of
Eq.(\ref{ladder2}) to the orthogonality conditions (\ref{orthog:short}). This
gives a system of two linear inhomogeneous equations: 
\begin{eqnarray}
a\omega \langle \langle \epsilon _{p}^{2}\rangle \rangle -b\frac{k^{2}}{3}\langle \langle \epsilon _{p}p^{2}\rangle \rangle  & = & 0,\label{abeq1} \\
-a\langle \langle p^{2}\epsilon _{p}\rangle \rangle +b\omega \langle \langle p^{2}\rangle \rangle  & = & -\langle \langle p^{2}\rangle \rangle .\label{abeq2} 
\end{eqnarray}

Eqs.~(\ref{abeq1}) and (\ref{abeq2}) describe the propagation of fluctuations
in the hydrodynamic thermal cloud. Without the rhs, the solutions of these equations
are free oscillations of a gas of excitations (first sound). They are
characterized by the dispersion relation \( \tilde{\omega }_{k}^{2}=k^{2}u^{2}, \)
where the velocity \( u \) of first sound is given by 
\[
u^{2}=\frac{\langle \langle \epsilon _{p}p^{2}\rangle \rangle ^{2}}{3\langle \langle \epsilon _{p}^{2}\rangle \rangle \langle \langle p^{2}\rangle \rangle }.\]
 In the temperature domain \( T\gg \mu  \) we have \( u^{2}=0.86T \).

For finding the damping rates of first sound from Eq.(\ref{losses:delta})
we are interested in frequencies \( \omega \approx \omega _{k}=c_{S}k \) which
at \( T\gg \mu  \) are much smaller than \( \tilde{\omega }_{k}=uk \). For
these \( \omega  \) the solutions of inhomogeneous Eqs.(\ref{abeq1}),(\ref{abeq2})
are 
\begin{equation}
\label{acousticab}
a=\frac{2}{3u^{2}};\, \, \, \, \, b=\frac{\omega }{u^{2}k^{2}},
\end{equation}
 and from Eq.(\ref{alpha:zero}) one gets a real response function. Thus, in
the acoustic approach, Eq.(\ref{losses:delta}) gives zero damping rate.

The damping of second sound appears in the so called viscous approximation
accounting for the first order correction \( \alpha _{{\textbf {p}}}^{(1)} \).
The latter is a solution of Eq.(\ref{ladder2}), with \( i=0 \) and \( \alpha ^{(0)}_{{\textbf {p}}} \)
from Eqs.~(\ref{alpha:zero}), (\ref{acousticab}): 
\[
{\textbf {kp}}(\omega b-a\epsilon _{p}+1)-\sum _{i,j}bk_{i}k_{j}(p_{i}p_{j}-\frac{\delta _{ij}}{3}p^{2})=i\tilde{I}_{{\textbf {p}}}[\alpha ^{(1)}_{{\textbf {p}}}].\]
 As the collisional integral vanishes for \( \alpha _{{\textbf {p}}}^{(1)}\! \propto \! \epsilon _{p} \)
and \( \alpha _{{\textbf {p}}}^{(1)}\! \propto \! {\textbf {kp}} \), the solution
of this equation can be written as 
\[
\alpha _{{\textbf {p}}}^{(1)}=\delta \alpha _{{\textbf {p}}}+\delta a\epsilon _{p}+\delta b({\textbf {kp}}).\]
 The quantity \( \delta \alpha _{{\textbf {p}}} \) can be represented in the
form 
\[
\delta \alpha _{{\textbf {p}}}=i({\textbf {kp}})\phi _{2}+ib\phi _{1}\sum _{i,j}k_{i}k_{j}(p_{i}p_{j}-\frac{\delta _{ij}}{3}p^{2}),\]
 where \( \phi _{1} \) and \( \phi _{2} \) are solutions of the equations
\begin{eqnarray}
\tilde{I}_{{\textbf {p}}}[(p_{i}p_{j}-\frac{\delta _{ij}}{3}p^{2})\phi _{1}] & = & (p_{i}p_{j}-\frac{\delta _{ij}}{3}p^{2}),\label{inversion:f1} \\
\tilde{I}_{{\textbf {p}}}({\textbf {p}}\phi _{2}) & = & {\textbf {p}}\left( 1-\frac{2\epsilon _{p}}{3u^{2}}\right) .\label{inversion:f2} 
\end{eqnarray}
 The functions \( \delta a \) and \( \delta b \) are obtained from the requirement
that both sides of Eq.(\ref{ladder2}) (with \( i=1 \)) are orthogonal to the
integrals of motion. The corresponding system of equations is similar to Eqs.(\ref{abeq1}),(\ref{abeq2})
and gives 
\begin{equation}
\label{a}
\! \! \delta a(\omega _{k},\! {\textbf {k}})\! =\! \frac{i\omega _{k}}{6u^{2}}\! \left( \! \frac{\langle \langle p^{4}\phi _{2}\rangle \rangle }{\langle \langle p^{4}\rangle \rangle }\! -\! 4\frac{\langle \langle p^{2}\phi _{2}\rangle \rangle }{\langle \langle p^{2}\rangle \rangle }\! -\! \frac{2}{5}\frac{\langle \langle p^{4}\phi _{1}\rangle \rangle }{\langle \langle p^{4}\rangle \rangle }\! \right) \! .\! \! 
\end{equation}
 Both \( \delta a,\delta b \) and \( \phi _{1},\phi _{2} \) turn out to be
independent of the direction of the vector \( {\textbf {p}} \). Therefore,
one easily finds that only the term \( \delta a\epsilon _{p} \) contributes
to the quantity \( \langle \langle \alpha _{{\textbf {p}}}^{(1)}\rangle \rangle  \).
Then, using Eq.(\ref{losses:delta}), for the damping rate of second sound
we obtain \( \Gamma _{k}=\omega _{k}g{\textrm{Im}}\langle \langle \delta a(\omega _{k},{\textbf {k}})\epsilon _{p}\rangle \rangle  \).
To calculate the averages containing \( \phi _{1} \) and \( \phi _{2} \) in
Eq.(\ref{a}), we invert numerically the collisional integral in Eqs.~(\ref{inversion:f1})
and (\ref{inversion:f2}). Then, at temperatures \( T\gg \mu =n_{0}g \) we
find 
\begin{equation}
\label{gamma:hom}
\Gamma _{k}\approx 4.8\frac{\omega _{k}^{2}}{n_{0}g}.
\end{equation}

The main contribution to \( \delta a(\omega _{k},{\textbf {k}}) \) in Eq.(\ref{a})
and to the damping rate \( \Gamma _{k} \) (\ref{gamma:hom}) is provided by
thermal excitations with energies \( \sim T \). As the characteristic relaxation
time for these single-particle excitations is \( \tau _{R}\sim (n_{0}\sigma v_{T})^{-1} \),
the criterion of the hydrodynamic regime for the damping of phonons with frequency
\( \omega _{k} \) takes the form 
\begin{equation}
\label{hydcr}
\omega _{k}\ll n_{0}\sigma v_{T}.
\end{equation}

The hydrodynamic criterion (\ref{hydcr}) is not exactly opposite to the condition of the collisionless regime,
\( \omega _{k}\gg\tilde{\tau}_{R}^{-1}\sim T(n_{0}a^{3})^{1/2} \), found in
\cite{fedichev:twopapers}.
In the latter case the collisional integral in Eq.(\ref{kineq:alpha})
is small and \( \alpha _{{\textbf {p}}}=-{\textbf {kp}}/(\omega -{\textbf {kp}}) \).
Then, Eq.(\ref{losses:delta}) gives the damping rate 
\begin{equation}
\label{landaudamp}
\Gamma ^{(L)}_{k}\sim \omega _{k}(n_{0}a^{3})^{1/2}\frac{T}{n_{0}g}
\end{equation}
which, except for the numerical coefficient, coincides with the rate of Landau
damping in the collisionless theory \cite{fedichev:twopapers}. The main contribution
to \( \Gamma _{k}^{(L)} \) (\ref{landaudamp}) comes from the interaction of
the oscillating condensate with thermal excitations which have energies \( \sim n_{0}g \)
\cite{com2} and are characterized by the relaxation rate \( \tilde{\tau }_{R}^{-1} \)
\cite{fedichev:twopapers}. For \( T\gg n_{0}g \) it exceeds the relaxation
rate \( \tau _{R}^{-1} \) of atoms with energies \( \sim T \): 
\( \tilde{\tau }_{R}/\tau _{R}\sim (n_{0}g/T)^{1/2}\ll 1 \).
Hence, there is a range of frequencies, where 
\( \tau _{R}^{-1}\alt \omega _{k}\alt \tilde{\tau }_{R}^{-1} \)
and neither the hydrodynamic nor the collisionless approach is valid.

The comparison of Eq.(\ref{gamma:hom}) with Eq.(\ref{landaudamp}) shows that
the ratio of the hydrodynamic to collisionless damping rate is \( \Gamma _{k}/\Gamma _{k}^{(L)}\sim \omega _{k}\tilde{\tau }_{R} \),
i.e. in the hydrodynamic regime the damping is much slower than one would estimate
with a collisionless approach.

We now calculate the hydrodynamic damping of the lowest excitations of a harmonically trapped
Thomas-Fermi condensate. In this case, for the density fluctuations we have 
(see \cite{Pitaevskii:review}) 
\begin{equation}
\label{deltan1}
\! \delta n_{0}(t,{\textbf {r}})=i\sum _{\nu }\left( \frac{\omega _{\nu }}{2n_{0}({\textbf {r}}}\right) ^{1/2}\! \! \! W_{\nu }({\textbf {r}})\{c_{\nu }e^{-i\omega _{\nu }t}-c.c.\}.
\end{equation}
 For a low-energy (\( \omega _{\nu }\ll \mu  \)) excitation with a set of quantum
numbers \( \nu  \), the function \( W_{\nu } \) is a polynomial of the coordinate
variables \( r_{i} \). The condensate density has the well-known parabolic
shape \( n_{0}({\textbf {r}})=n_{0m}(1-\sum _{i}r_{i}^{2}/l_{i}^{2}) \), with
the maximum value \( n_{0m}=n_{0}(0)=\mu /g \) and the size \( l_{i}=(2\mu /\omega _{i}^{2})^{1/2} \)
in the \( i \)-th direction.

The wavelength of the lowest excitations is of order the spatial size of the
condensate, which in the hydrodynamic regime greatly exceeds the mean free path
of thermal excitations. Therefore, we will use a local density approach assuming
that in the vicinity of a point \( {\textbf {r}} \) the generalized susceptibility
is the same as in a homogeneous gas with the condensate density \( n_{0}({\textbf {r}}) \).
This is justified by a local spatial character of the collisional integral.
As Eq.(\ref{deltan1}) only contains an extra factor \( W_{\nu }({\textbf {r}}) \)
compared to Eq.(\ref{deltan}), the damping rate is given by \( \Gamma _{\nu }=\omega _{\nu }g\int d{\textbf {r}}|W_{\nu }({\textbf {r}})|^{2}{\textrm{Im}}\langle \langle \alpha _{{\textbf {p}}}(\omega _{\nu },{\textbf {k}},n_{0}({\textbf {r}}))\rangle \rangle  \)
and reduces to 
\begin{equation}
\label{G}
\Gamma _{\nu }=\int d{\textbf {r}}|W_{\nu }({\textbf {r}})|^{2}\Gamma _{\nu h}({\textbf {r}}).
\end{equation}
The quantity \( \Gamma _{\nu h}({\textbf {r}}) \) is the damping rate of first sound phonons with frequency
\( \omega_{\nu} \) in a homogeneous gas with the condensate density \( n_0({\bf r}) \). This quantity is given by Eq.(\ref{gamma:hom})
with \( n_{0}=n_{0}({\textbf {r}}) \), and the integral in Eq.(\ref{G}) is
logarithmically divergent at the condensate boundary. The divergency originates
from our assumption that \( n_{0} \) is much larger than the density of non-condensed
atoms \( n' \), which is violated near the Thomas-Fermi border. Hence, we have
to cut-off the integration at distances where \( n_{0}({\textbf {r}})\sim n' \).
Then, expressing \( W_{\nu } \) in terms of the radial variable \( y=(\sum _{i}r_{i}^{2}/l_{i}^{2})^{1/2} \)
and angle variables \( \theta ,\varphi  \), with logarithmic accuracy we obtain
\begin{equation}
\label{Gf}
\Gamma _{\nu }=2.4\gamma _{\nu }\frac{\omega _{\nu }^{2}}{n_{0m}g}\log {\left( \frac{n_{0m}}{n'}\right) },
\end{equation}
 where \( \gamma _{\nu }=\int d\Omega |W_{\nu }(y=1,\theta ,\varphi )|^{2} \).

Our approach describes the damping of oscillations of a trapped condensate,
assuming that the latter was set into motion while not directly affecting the
thermal cloud (or vice versa). Using the polynomials \( W_{\nu } \) from \cite{Ohberg},
we calculated the coefficients \( \gamma _{\nu } \) for several particular
cases. In spherical traps we find \( \gamma _{\nu }=(4n+2l+3) \), where \( n \)
is the radial quantum number and \( l \) the orbital angular momentum. For
the so called surface excitations (\( n=0 \)) this result remains valid in
cylindrical traps, and Eq.(\ref{Gf}) indicates a significant increase of the
damping rate with increasing \( l \). However, it is less dramatic than observed
at MIT \cite{MIT:surfacemodes} in the collisionless regime. For the dipole
axial oscillations of the condensate we obtain \( \gamma _{\nu }=5 \). Then,
for the conditions of the MIT experiment \cite{MIT:shifts} (\( \omega _{z}=18 \)Hz,
\( T\approx 1\mu  \)K, \( \mu \approx 200 \)nK), Eq.(\ref{Gf}) gives \( \Gamma _{\nu }\approx 10 \)
s\( ^{-1} \). This agrees with the experiment which was entering the hydrodynamic
regime for the axial motion.

We believe that our approach can be further developed to describe damping phenomena
in various intermediate regimes which one can be meet in trapped gases.

We acknowledge fruitful discussions with M.A. Baranov, A. Griffin, and J.T.
M. Walraven. This work was financially supported by the Stichting voor Fundamenteel
Onderzoek der Materie (FOM), by INTAS, and by the Russian Foundation for Basic
studies.  \bibliographystyle{prsty}
\bibliography{myDb}

\begin{thebibliography}{10}
\bibitem{com1}
Our definition of the susceptibility differs by a factor of \( (\partial
  f_0/\partial \epsilon_p) \) from the usual definition in \cite{LL:volX}.
\bibitem{com2}
At these energies the excitations have a collective character, and for this
  reason Eq.(\ref{landaudamp}) based on the single-particle approach does not
  recover a correct numerical coefficient.
\end{thebibliography}

\end{document}